# Marx’s two aggregate invariants and the solution of the transformation problem in a Morishima-type formalism

Norbert Ankri[1*] and Païkan Marcaggi[2]

[1] INSERM - Aix-Marseille Université, UMR1072, Marseille, France

[2] Muséum National d’Histoire Naturelle, CNRS UMR7196, INSERM U1154, Paris, France

## Abstract

The unit value of a commodity, as correctly determined by Michio Morishima’s method and its variations, is the sum of the labour-values of its inputs and the direct labour required for its production. However, commodities are sold at their prices of production only when they meet effective social demand and are thereby validated as realised social value. Price determination involves the entire economy and the interconnections among its different branches. This condition gives full meaning to Marx’s fundamental equalities, which derive from the law of value and function as invariants governing the economy as a whole. These equalities are necessary, and, once combined with the relevant reproduction constraints and a specified wage basket, sufficient to determine the admissible transformation in the simple reproduction cases examined here. We demonstrate this by starting from Morishima’s formalism and returning to a formalism closer to that used by Marx.

[*] Correspondence: norbert.ankri@univ-amu.fr

## 1. Introduction

In Capital Volume 1 (Marx, 1976[1867]), Marx explains that the prices of commodities oscillate around an average price, namely their value. In Capital Volume 3 (Marx, 1981[1894]), the notion of the price of production of a commodity is introduced: it is now around this price of production, and no longer directly around value, that market prices tend to oscillate. At first sight, the theory of value may therefore seem to lose coherence. It becomes necessary to show how values are related to prices of production, that is, what operation enables one to transform the former into the latter. To illustrate the difference between values and prices of production, Marx considers a model composed of five branches, in each of which an identical capital of 100 monetary units is invested. In this example, the inputs are not transformed into prices of production. Marx uses it to support a specific aspect of his theory: how surplus-value is drawn preferentially towards the most capital-intensive sectors, and how this movement is masked in the real world, where values are not directly observable. In this example, however, there is no question of simple reproduction and no explicit interconnection among the different branches. These issues are addressed by Marx's law of value, which implies two fundamental aggregate equalities: the equality between the sum of profits and the sum of surplus-values, and the equality between the total value of the realised social product and its total price of production. These equalities, commonly referred to as invariants, make sense only on the scale of the economy as a whole. At this aggregate level, the interconnections among branches mean that the amount of capital allocated to them cannot be chosen arbitrarily; it must comply with effective social demand[1]. Marx's fundamental equalities are analogous to conservation laws: they govern the connection between values and prices while allowing the redistribution of surplus-value among branches through the equalisation of the profit rate.

[1] The notion of effective social demand used here is related to Vincent Laure van Bambeke's emphasis on 'besoins sociaux solvables' (Laure van Bambeke, 2013, 2021; see also Harribey, 2022). However, the present approach differs from Laure van Bambeke's method and from his interpretation of Marx's aggregate equalities.

An explanation of how values and prices of production are connected is provided in Capital Volume 3 (Marx, 1981[1894]). However, the underlying mathematical method for transforming values into prices of production is not explicitly supplied. This incompleteness does not appear to have been a major concern for Marx, as if the missing mathematics were a secondary and solvable problem. Since then, however, the so-called "transformation problem" has generated a long-lasting controversy, and different formalisms have been developed to address it, with Morishima's input-output formalism gaining influence by providing a straightforward linear-algebraic method (Morishima, 1973, 1974).

Here, we identify a limitation in certain uses of Morishima-like formalisms, namely when the technical system, together with a given net-output vector, is treated as sufficient to determine the realised price–quantity configuration[2]. The limitation lies not in the computation of labour-values itself, but in treating the realised configuration as if it could be determined without introducing effective social demand as an independent constraint. The technological possibilities of a given period are indeed essential constraints, but they are not the only ones. A second constraint is social and economic: in a capitalist market, a commodity can be sold at a price incorporating the current average rate of profit only if it meets effective social demand. This condition gives concrete content to Marx's aggregate equalities. It follows that the condition of simple reproduction cannot be validated by considering only the technical conditions of production. Both values and realised demand must first be considered. In the stationary benchmark examined here, this means that the produced quantities are assumed to coincide with the quantities socially validated through sale. In order to comply with Marx's fundamental equalities, we show how a transition can be made from Morishima's common input-output formalism, as clearly presented by Basu (2021), to a formulation closer to Marx's original categories. We first express the rationale for this transition using Basu's three-branch example, where two branches produce commodities that can be consumed by both industries and workers, while the remaining branch produces a commodity used by industries only. A generalization to n branches is provided in the appendix.

[2] By "price–quantity configuration" we mean the pair formed by the production-price vector and the sectoral quantities actually realised in social reproduction. This question arises whenever sectoral quantities are treated independently of effective demand.

## 2. Link with the two-sector law

The two-sector model developed in Ankri (in press) establishes the following result: proportional consistency between production and effective social demand fixes the capital allocation compatible with value conservation, profit realisation and uniform profitability. In that model, branch 1 produces means of production and branch 2 produces wage goods. Let $W_1$ and $W_2$ denote the total values of the outputs of branches 1 and 2, respectively. Let $C=C_1+C_2$ denote the total means of production required by both branches, and let $V=V_1+V_2$ denote the total wage goods required by workers in both branches. Proportional consistency between production and effective social demand requires (Ankri, in press):

$$C / V = W_1 / W_2.$$

This condition is not a merely formal identity. It expresses the requirement that the output of the branch producing means of production and the output of the branch producing wage goods be proportionally consistent with the social demand generated by the economy itself. The three-branch case developed below should be read as a generalization of the same logic. Adding a third commodity relaxes the rigid proportionality of the two-branch model. Ankri (in press) isolates the elementary proportionality condition; the present paper shows how the same two aggregate conservation principles operate in a Morishima-type transformation framework built on Leontief input-output relations, where the price vector, the wage basket and the allocation of capital interact more flexibly.

## 3. The three-branch input-output economy

Consider three branches producing bread, iron and meat. Let A be a non-negative productive input-output matrix.

$$A = [a_{ij}]_{i,j=1,2,3}.$$

The coefficient $a_{ij}$ denotes the physical amount of commodity i required to produce one unit of commodity j. For comparability with a previous numerical example (Basu, 2021), the labels bread, iron and meat are retained. They should not be read as an empirical input-output table for actual agriculture, metallurgy or meat production. They denote three commodities in an abstract linear reproduction system. Bread and meat are wage-goods entering workers' consumption, while iron is treated as a production good. In the numerical example, workers consume bread and meat; iron is used only as a means of production.

Let $\ell = (\ell_1, \ell_2, \ell_3)$ be the row vector of direct labour requirements, with $\ell_j$ denoting the direct labour required to produce one unit of commodity j. Let $\Lambda = (\Lambda_1, \Lambda_2, \Lambda_3)$ be the row vector of unit labour-values. The value equation is given by

$$\Lambda = \Lambda A + \ell$$

$$\Lambda(I - A) = \ell$$

Provided that the Hawkins-Simon productivity condition (Hawkins and Simon, 1949) is satisfied, so that the Leontief inverse $(I - A)^{-1}$ exists and is non-negative, $\Lambda$ is given by

$$\Lambda = \ell(I - A)^{-1} \quad (1)$$

Let $v = (v_1, v_2, v_3)^T$ be the real wage basket paid per unit of labour. The value of v is $\Lambda v$.

Since one hour of labour creates one unit of new value, the rate of exploitation e is

$$e = (1 - \Lambda v) / (\Lambda v). \quad (2)$$

We define α as the value share of bread in the wage basket:

$$\alpha = \Lambda_1 v_1 / (\Lambda_1 v_1 + \Lambda_2 v_2 + \Lambda_3 v_3) = \Lambda_1 v_1 / \Lambda v. \quad (3)$$

In this model, workers' consumption consists only of bread and meat, while iron is not part of the wage basket: $v_2 = 0$ and equation (3) reduces to $\alpha = \Lambda_1 v_1 / (\Lambda_1 v_1 + \Lambda_3 v_3)$. Thus α = 1 corresponds to a wage basket composed only of bread, whereas α = 0 corresponds to a wage basket composed only of meat.

This part of Morishima's formalism, summarized in equations (1) and (2), is retained. The point of disagreement with purely technical solutions is not the computation of Λ, but the idea that Λ and A alone could determine production prices without specifying the macroeconomic realisation constraints. By purely technical solutions, we mean approaches that derive prices and quantities from A, ℓ and a given net-output vector, without treating effective social demand and market validation as independent constraints. By macroeconomic realisation constraints, we mean the aggregate conditions under which produced commodities are socially validated as realised value. To apply aggregate conditions, we translated Morishima's formalism into Marxian categories.

## 4. Translation into Marxian categories

For each branch j, the total value of output $W_j$ can be decomposed into constant circulating capital $C_j$, variable capital $V_j$, and surplus-value $S_j$. Here $C_j$ denotes the value of the means of production used up in branch j, $V_j$ denotes the value of the wage goods consumed by the workers employed in branch j, $K_j$ denotes the capital assigned to branch j ($K_j = C_j + V_j$), $S_j$ denotes the surplus-value produced in branch j. If $q_j$ denotes the gross output of commodity j, then the value of commodity i used as circulating input in branch j is $\Lambda_i a_{ij} q_j$. The value of commodity i entering the wage basket consumed by workers employed in branch j is $\Lambda_i v_i \ell_j q_j$. Surplus-value is the excess of newly added labour-value over the value of labour-power:

$$C_j = \Sigma_i \Lambda_i a_{ij} q_j \quad (4a)$$

$$V_j = \Sigma_i \Lambda_i v_i \ell_j q_j = (\Lambda v)\ell_j q_j \quad (4b)$$

$$S_j = \ell_j q_j - V_j = eV_j \quad (4c)$$

$$W_j = C_j + V_j + S_j = K_j + S_j \quad (4d)$$

Lower-case letters denote unit magnitudes (apart from q); upper-case letters denote aggregate magnitudes after the scale of production and capital allocation have been chosen. Since branch j produces commodity j, the $W_j$ magnitude is also given by

$$W_j=\Lambda_j q_j$$

The crucial methodological point is that ($q_1$, $q_2$, $q_3$), or equivalently ($K_1$, $K_2$, $K_3$), cannot be chosen arbitrarily. Commodities that are physically producible do not thereby constitute realised value. They count as realised social value only insofar as they are validated by sale. The reproduction problem is therefore simultaneously technical and social.

## 5. Production prices and the uniform profit rate

The following shows how we can express the transformation in a notation closer to Marxian categories. In the three-branch model used here, branch 1 produces bread, branch 2 produces iron, and branch 3 produces meat. The technical input matrix is denoted by $A = [a_{ij}]_{i,j=1,2,3}$, the direct-labour vector by $\ell = (\ell_1, \ell_2, \ell_3)$, and the unit labour-value vector by $\Lambda = (\Lambda_1, \Lambda_2, \Lambda_3)$. Workers consume bread and meat, while iron is not part of the wage basket. Thus the physical wage basket is $v = (v_1, 0, v_3)^T$.

Real wage being represented by the physical basket v, the relevant input matrix for the price system is the augmented matrix $\tilde{A}$, which includes both productive inputs and the commodity composition of wages:

$$\tilde{A} = A + v\ell. \quad (5.1)$$

Here $v\ell$ is the matrix with entries $v_i\ell_j$, so that

$$\tilde{A} = [\tilde{a}_{ij}]_{i,j=1,2,3}, \quad \tilde{a}_{ij} = a_{ij} + v_i\ell_j.$$

In the augmented matrix $\tilde{A}$, the added term in position (i,j), namely $v_i\ell_j$, represents the wage-good requirement associated with the labour employed in branch j, not an additional technical input coefficient.

For each branch j = 1, 2, 3, the lowercase magnitudes $b_j$, $f_j$, and $m_j$ denote, respectively, the value of bread, iron[3], and meat required to produce one unit of commodity j, once the wage basket has been incorporated into $\tilde{A}$. Note that $b_j$, $f_j$, and $m_j$ are value magnitudes, not physical input coefficients.

The value of bread required in branch j is decomposed into a circulating-capital component and a variable-capital component:

$$b_j^c = \Lambda_1 a_{1j}, \quad b_j^V = \Lambda_1 v_1 \ell_j, \quad b_j = b_j^c + b_j^V = \Lambda_1 \tilde{a}_{1j} \quad (5.2)$$

The superscripts c and v indicate, respectively, the part belonging to circulating constant capital and the part belonging to variable capital. Since iron is not part of the workers' wage basket, its value component is simply:

$$f_j = \Lambda_2 a_{2j} + 0 = \Lambda_2 \tilde{a}_{2j} \quad (5.3)$$

The value of meat required in branch j is similarly decomposed as:

---

[3] Iron is a chemical element named Fe from the Latin *ferrum*. We use letter f to designate it.

$$m_j^c = \Lambda_3 a_{3j}, \quad m_j^V = \Lambda_3 v_3 \ell_j, \quad m_j = m_j^c + m_j^V = \Lambda_3 \tilde{a}_{3j} \qquad (5.4)$$

From equation (2), we have

$$\Lambda v = \Lambda_1 v_1 + \Lambda_3 v_3 = 1/(1+e)$$

The value share of bread in the wage basket is

$$\alpha = \Lambda_1 v_1/\Lambda v = \Lambda_1 v_1/(\Lambda_1 v_1 + \Lambda_3 v_3)$$

Consequently,

$$b_j^V = \alpha \ell_j/(1+e), \quad m_j^V = (1-\alpha)\ell_j/(1+e).$$

The surplus-value produced per unit of commodity j is denoted by $s_j$:

$$s_j = \ell_j - (b_j^V + m_j^V) = e\ell_j/(1+e) \qquad (5.5)$$

The unit value of the commodity produced by branch j is therefore:

$$\Lambda_j = b_j + f_j + m_j + s_j.$$

The value of the capital advanced per unit of commodity j is:

$$k_j = b_j + f_j + m_j$$

The transformation of values into prices is represented by the coefficients $x_i$. For commodity i, the production price $p_i$ is defined by:

$$p_i = \Lambda_i x_i,$$

If $x_i = 1$, the production price of commodity i is equal to its labour-value. If $x_i > 1$, it is above its labour-value; if $x_i < 1$, it is below its labour-value.

For one unit of output in branch j, the capital advanced expressed in production prices is:

$$k_j^p = x_1 b_j + x_2 f_j + x_3 m_j.$$

The profit realised on one unit of output in branch j is denoted by $\pi_j$:

$$\pi_j = \Lambda_j x_j - k_j{}^p.$$

At this level, $\pi_j$ and $s_j$ are unit magnitudes. Their aggregate counterparts are $\Pi_j = q_j\pi_j$ and $S_j = q_j s_j$. Thus $\Pi_j$ denotes profit, whereas $S_j$ denotes surplus-value. These magnitudes must not be confused.

Uniform profitability requires that the rate of profit be the same in the three branches:

$$\pi_j/k_j{}^p = r, \quad j = 1, 2, 3.$$

Equivalently, for each branch j:

$$\Lambda_j x_j = (1+r)k_j{}^p.$$

The system of price equations can therefore be written explicitly as:

$$\Lambda_1 x_1 = (1+r)(b_1 x_1 + f_1 x_2 + m_1 x_3),$$

$$\Lambda_2 x_2 = (1+r)(b_2 x_1 + f_2 x_2 + m_2 x_3),$$

$$\Lambda_3 x_3 = (1+r)(b_3 x_1 + f_3 x_2 + m_3 x_3). \quad (5.6)$$

The subscript on $x_i$ refers to the commodity whose value is transformed into a price: $x_1$ transforms bread, $x_2$ transforms iron, and $x_3$ transforms meat. By contrast, the subscript on $b_j$, $f_j$, and $m_j$ refers to the branch in which these input values are required.

To express the system in linear algebra notation, we define the matrix

$$Z = [z_{jk}]_{j,k=1,2,3}$$

by

$$z_{j1} = b_j/\Lambda_j, \quad z_{j2} = f_j/\Lambda_j, \quad z_{j3} = m_j/\Lambda_j.$$

With the column-vector convention $x = (x_1, x_2, x_3)^T$, the system of price equations can be written as the eigenvalue problem

$$Zx = (1/(1+r))x \quad (5.7)$$

or

$$Zx = \lambda x$$

where λ denotes the relevant eigenvalue of Z

$$r = 1/\lambda - 1 \qquad (5.8)$$

Under the standard productivity and irreducibility assumptions, the economically relevant relative solution is the Perron–Frobenius eigenvector associated with the maximal eigenvalue of Z. Let the eigenvector normalised to unit length be denoted by

$$x^* = (x_1^*, x_2^*, x_3^*)^T$$

It determines the transformation coefficients only up to a common scalar factor γ, the economically relevant coefficients being written:

$$x_j = \gamma\, x_j^*, \quad j = 1, 2, 3 \quad (5.9)$$

For each branch j, the aggregate capital advanced in values is

$$K_j = q_j\, k_j$$

The aggregate value of the output of branch j is therefore

$$W_j = \Lambda_j q_j = \Lambda_j K_j / k_j.$$

Marx's aggregate equality between total production price and total value of the realised social product can be written:

$$\Sigma_j\, p_j q_j = \Sigma_j\, \Lambda_j q_j.$$

since

$$p_j = \Lambda_j x_j = \gamma \Lambda_j x_j^*,$$

we have

$$\gamma \Sigma_j\, x_j^* W_j = \Sigma_j\, W_j,$$

and therefore

$$\gamma = (\Sigma_j W_j) / (\Sigma_j x_j^* W_j)$$

$$= (\Sigma_j \Lambda_j K_j/k_j) / (\Sigma_j x_j^* \Lambda_j K_j/k_j) \qquad (5.10)$$

Thus the Perron–Frobenius eigenvector fixes the relative transformation coefficients, while Marx's aggregate equality fixes their economically relevant scale once the admissible capital allocation has been determined. Monetary units are treated here as an ideal expression of labour-time: one monetary unit represents one hour of socially necessary labour. Unlike in the New Interpretation, money is not introduced as an additional closure condition of the system; real-currency issues are deliberately left outside the model. The corresponding total capital advanced in production prices is $K_j^p = q_j k_j^p$, while as stated above $S_j = q_j s_j$ and $\Pi_j = q_j \pi_j$.

Marx's two aggregate equalities can now be written in the notation used above. The first equality states that total profit must be equal to total surplus-value:

$$\Sigma_j \Pi_j = \Sigma_j S_j, \text{ equivalently } \Sigma_j q_j \pi_j = \Sigma_j q_j s_j \qquad (5.11)$$

The second equality states that the total production price of the realised social product must be equal to its total value:

$$\Sigma_j p_j q_j = \Sigma_j \Lambda_j q_j$$

This equality has already been used above to fix the scalar factor $\gamma$. Given the first equality, it can equivalently be written, in the present circulating-capital framework, as equality between total capital advanced in production prices and total capital advanced in values:

$$\Sigma_j K_j^p = \Sigma_j K_j, \text{ equivalently } \Sigma_j q_j k_j^p = \Sigma_j q_j k_j \qquad (5.12)$$

These two equalities are conservation constraints at the level of the economy as a whole. They do not apply branch by branch. They also imply that the allocation of capital across branches cannot be arbitrary: it must be compatible with the effective social demand that validates commodities as realised value.

## 6. Simple reproduction constraints and capital allocation

The uniform-profit-rate equations determine r and the relative transformation vector x*. Marx's second aggregate equality fixes the scalar γ, and hence the economically relevant transformation coefficients $x_j = \gamma x_j^*$, provided the admissible allocation of capital across branches is known. The latter must satisfy the relevant reproduction closure conditions.

Let $K_T$ denote the total capital advanced. The admissible allocation of capital is obtained by solving for $K_1$, $K_2$, and $K_3$ under the following three conditions. First,

$$K_1 + K_2 + K_3 = K_T \quad (6.1a)$$

Second, since $q_j = K_j/k_j$, total surplus-value can be written as

$$\Sigma_j S_j = \Sigma_j q_j s_j = \Sigma_j K_j (s_j/k_j).$$

Together with Marx's equality between total surplus-value and total profit, and hence with the conservation of total advanced capital, this gives

$$\Sigma_j K_j (s_j/k_j) = rK_T \quad (6.1b)$$

Third, the reproduction condition for iron is

$$q_2 \Lambda_2 = q_1 f_1 + q_2 f_2 + q_3 f_3.$$

Using $q_j = K_j/k_j$, this becomes

$$(K_1/k_1) f_1 + (K_2/k_2)(f_2 - \Lambda_2) + (K_3/k_3) f_3 = 0. \quad (6.1c)$$

Thus, for a chosen total capital $K_T$, the admissible capital allocation is determined by the system (6.1a)–(6.1c):

$$K_1 + K_2 + K_3 = K_T \quad (6.1a)$$

$$\Sigma_j K_j(s_j/k_j) = rK_T \quad (6.1b)$$

$$(K_1/k_1)f_1 + (K_2/k_2)(f_2 - \Lambda_2) + (K_3/k_3)f_3 = 0 \quad (6.1c)$$

Once $K_1$, $K_2$, and $K_3$ are known, the corresponding gross outputs are obtained from $q_j = K_j/k_j$, and the aggregate magnitudes reported in the tables follow.

In the circulating-capital case considered here, the profit of branch j is

$$\Pi_j = q_j[\Lambda_j x_j - (b_j x_1 + f_j x_2 + m_j x_3)]$$

while the surplus-value of branch j is

$$S_j = q_j s_j$$

The algebraic role of Marx's first aggregate equality can now be made explicit in the three-commodity notation. Equality between total profit and total surplus-value gives:

$$q_1[x_1(\Lambda_1 - b_1) - x_2 f_1 - x_3 m_1 - s_1] + q_2[x_2(\Lambda_2 - f_2) - x_1 b_2 - x_3 m_2 - s_2] + q_3[x_3(\Lambda_3 - m_3) - x_1 b_3 - x_2 f_3 - s_3] = 0. \quad (6.2)$$

Rearranging equation (6.2) by grouping terms by commodities rather than by branches, and considering the limiting case in which surplus-value is set to zero, gives:

$$[q_1\Lambda_1 - (q_1 b_1 + q_2 b_2 + q_3 b_3)]x_1 + [q_2\Lambda_2 - (q_1 f_1 + q_2 f_2 + q_3 f_3)]x_2 + [q_3\Lambda_3 - (q_1 m_1 + q_2 m_2 + q_3 m_3)]x_3 = 0. \quad (6.3)$$

The bracketed terms in equation (6.3) are the value expressions of the residual surplus outputs of bread, iron and meat after replacement of augmented input requirements. They therefore measure the difference between the output of each commodity and the socially validated requirements for that commodity. Under the assumption that effective social needs are satisfied, these terms are non-negative. Since the transformation coefficients $x_i$ are strictly positive and the bracketed residual-output terms are non-negative under the effective-demand assumption, equation (6.3) can hold in the zero-surplus limiting case only if each bracketed term vanishes. This determines the relative capital distribution; the absolute scale remains arbitrary until a normalisation, such as total realised value, is imposed.

Economically, a positive bracket would represent a residual surplus commodity; but when total surplus-value is zero, no residual surplus output can be validated. A negative bracket would mean an unsatisfied replacement or final-demand requirement; it is therefore excluded from the admissible reproduction configuration.

With positive surplus-value, the same reasoning no longer implies that all residual surplus outputs vanish. The corresponding residual outputs may be positive, since they represent surplus commodities validated by effective demand for non-wage final uses, such as capitalist consumption, accumulation, or other final outlets. In the example considered here, however, iron enters neither the wage basket nor capitalist final consumption. Its socially validated requirement is therefore limited to its productive use as an input as stated above:

$$q_2\Lambda_2 = q_1f_1 + q_2f_2 + q_3f_3. \qquad (6.4)$$

This reproduction condition, specific to the iron branch, constitutes the third equation of the capital-allocation system (6.1a)–(6.1c).

The unit capitals are already given by

$$k_j = b_j + f_j + m_j,$$

and the gross outputs are linked to the capital allocation by

$$q_j = K_j/k_j.$$

Thus, for a given total capital $K_T$, the admissible vector ($q_1$, $q_2$, $q_3$), equivalently the capital vector ($K_1$, $K_2$, $K_3$), is obtained from the system (6.1a)–(6.1c). Together with the previously determined uniform profit rate, transformation coefficients and wage basket, this system determines the relevant solution used in the numerical examples.

Thus, for simple reproduction, the admissible capital allocation is not obtained from the technical matrix alone. It is obtained only after Marx's aggregate equalities have been combined with the relevant reproduction conditions, such as equation (6.4) for iron in the present example.

Relative to the earlier version of this work (arXiv:2303.11471v2), the claim is therefore corrected as follows: Marx's aggregate equalities are necessary for the transformation, and, when combined with the reproduction constraints exemplified by equations (6.3) or (6.4) and a specified wage basket, they are sufficient to determine the relevant prices and feasible capital allocations in the simple reproduction cases examined here. Taken by themselves, however, Marx's aggregate equalities restrict but do not uniquely determine the admissible allocation of total capital across sectors; they must be supplemented by the reproduction constraints and by the specification of the wage basket.

## 7. Physical quantities and the net-output vector

The preceding equations determine values, prices, profit rates and admissible capital allocations. They also determine the corresponding physical quantities. This point is important because the transformation problem cannot be solved by arbitrarily choosing a physical net-output vector independently of effective social demand.

Let A be the technical input-output matrix. It records the physical commodities directly required as productive inputs. In the standard Leontief definition used below (Leontief, 1986), the net-output vector y is computed after replacement of these productive inputs only (so y includes the wage basket).

In contrast, the augmented matrix $\tilde{A}$ incorporates the wage basket:

$$\tilde{A} = [\tilde{a}_{ij}]_{i,j=1,2,3} = A + v\ell$$

Here v is the physical wage basket paid per unit of labour and $\ell$ is the row vector of direct labour requirements. Thus the entries of $\tilde{A}$ are

$$\tilde{a}_{ij} = a_{ij} + v_i\ell_j.$$

In the bread-iron-meat example, workers consume bread and meat, while iron is not part of the wage basket. Hence only the bread and meat rows of the initial matrix are increased.

If the wage goods are initially expressed in value terms rather than in physical quantities, they can be converted back into physical quantities by division by their corresponding unit values. Let $b_j^V$ denote the value of the bread component of the wage basket required per unit of output in branch j, and let $m_j^V$ denote the corresponding value of the meat component. Then

$$\tilde{a}_{1j} = (a_{1j}\Lambda_1 + b_j^V)/\Lambda_1 = a_{1j} + b_j^V/\Lambda_1,$$

$$\tilde{a}_{2j} = a_{2j},$$

$$\tilde{a}_{3j} = (a_{3j}\Lambda_3 + m_j^V)/\Lambda_3 = a_{3j} + m_j^V/\Lambda_3.$$

If wages are zero, $\tilde{A} = A$.

In what follows, no separate labour-value vector is required. The labour-values used throughout remain $\Lambda = \ell(I - A)^{-1}$.

The vector $\Lambda$ gives the labour-values of commodities produced with the technical matrix A. By contrast, $\tilde{A}$ is not used to redefine labour-values. It is used to express the price system and the reproduction constraints once the wage basket has been incorporated into the input requirements.

Let $q = (q_1, q_2, q_3)^T$ be the gross-output vector, where $q_j$ is the physical quantity of commodity j produced. Let $y = (y_1, y_2, y_3)^T$ be the net-output vector, that is, the quantities remaining after replacement of the technical input requirements recorded in A. Then

$$q = Aq + y \qquad (7a)$$

$$y = (I - A)q \qquad (7b)$$

and, provided I − A is invertible,

$$q = (I - A)^{-1}y \qquad (7c)$$

Here, the vector y, the physical net-output vector by commodity, is not introduced as an exogenous datum. In the present case, iron is neither part of the wage basket nor an object of capitalist or unproductive consumption. Simple reproduction therefore requires:

$$y_2 = 0 \qquad (7d)$$

Bread and meat, by contrast, may have positive net outputs, since they can enter socially validated final consumption, including wage consumption and, when surplus-value is positive, non-wage final demand.

At the aggregate level, the value expression of the standard net product (y) corresponds to the total new value created by living labour, that is, (V+S). After subtraction of the wage-goods component, the residual surplus-output vector (h) corresponds, in value terms, to total surplus-value and, in price terms, to total profit. This correspondence does not hold branch by branch, because profits are realised on sales both as inputs to other branches and as final consumption, and because production prices redistribute surplus-value among branches.

In the numerical tables, once the aggregate value $W_j$ of branch j and the unit labour-value $\Lambda_j$ of its commodity are known, the corresponding gross output is obtained as:

$$q_j = W_j/\Lambda_j.$$

The net-output vector is then computed from equation (7b). It should therefore be read in the standard input-output sense: it includes wage goods as well as surplus commodities (absorbed only through capitalist consumption, accumulation, or other non-wage final outlets). In order to isolate this residual surplus, its vector is denoted

$$h = (I - \tilde{A})q = y - v(\ell q)$$

This residual-output vector can be treated as realised value only insofar as its components are validated by effective social demand. If additional quantities are produced beyond these socially validated outlets, they cannot be treated as realised value; they are unsold, devalued, or destroyed as value at the moment of market validation.

Since $\Lambda(I - A) = \ell$, it follows that $\Lambda(I - A)q = \ell q$, so $\Lambda y = \ell q$

It follows that

$\Lambda h = \Lambda y - \Lambda v(\ell q) = (1 - \Lambda v)\ell q = \Sigma_j S_j$

The condition (7d) has already been imposed at the level of capital allocation, through the reproduction condition for iron. It means that the whole output of the iron branch must exactly replace the iron used as input by the three branches.

By contrast, the net outputs of bread and meat need not vanish. They may correspond to wage goods and, when surplus-value is positive, to surplus commodities validated by effective demand. The essential point is that the vector y cannot be freely chosen as an arbitrary physical datum. Its components are linked, on the one hand, by the technical requirements of production and, on the other hand, by the allocation of capital according to effective social demand.

This is why, in the present approach, the net-output vector is not introduced as an exogenous datum. In a capitalist economy, the net product is not decreed independently of the market. It must be compatible with the value structure of production, with Marx's aggregate conservation equalities and with the social validation of commodities through sale.

Finally, the composition of the wage basket matters. If two wage baskets have the same total value $\Lambda v$, they imply the same rate of exploitation. However, they may still modify the augmented matrix $\tilde{A} = A + v\ell$, and therefore the intersectoral structure of prices, quantities and the uniform profit rate. This is why the value of labour-power determines exploitation, while the concrete commodity composition of workers' consumption can still affect the reproduction structure of the system.

## 8. Numerical results

To illustrate the method, we use the same three-branch numerical example as Basu (2021), except for the net-output vector y, which cannot be chosen arbitrarily in the present approach. The technical coefficients and direct-labour requirements are given by:

$$A = \begin{pmatrix} 186/450 & 54/21 & 30/60 \\ 12/450 & 6/21 & 3/60 \\ 9/450 & 6/21 & 15/60 \end{pmatrix}$$

and $\ell = (18/450, 12/21, 30/60)$.

The first row of A gives the bread requirements, the second row the iron requirements, and the third row the meat requirements. The columns refer respectively to the bread, iron and meat branches. The vector $\ell$ gives the direct labour required to produce one unit of each commodity.

From equation (1), the corresponding unit labour-values are:

$\Lambda = \ell(I - A)^{-1} = (2/11, 20/11, 10/11)$,

| | $\Lambda$ |
|---|---|
| *Bread* | *0.181818182* |
| *Iron* | *1.818181818* |
| *Meat* | *0.909090909* |
| *Total* | *2.909090909* |

The numerical cases below differ only by the wage basket and by the corresponding rate of surplus-value. In each case, the augmented matrix $\tilde{A} = A + v\ell$ is constructed from the relevant wage basket, and the admissible production scale is then determined by Marx's two aggregate invariants together with the reproduction condition for iron, $y_2 = 0$, where applicable.

### 8.1 Reference wage case

$v = (2, 0, 1/6)^T$
From equation (3), $\alpha = 12/17$
From equation (2), $e = 16/17$

***Table 1. Unit value table*** (for the production of one unit of each commodity)

$b_j$, $f_j$ and $m_j$ and $s_j$ are obtained from equations (5.2), (5.3), (5.4) and (5.5), respectively.

| | b | f | m | s | $\Lambda$ |
|---|---|---|---|---|---|
| *Bread* | *0.089696970* | *0.048484848* | *0.024242424* | *0.019393939* | *0.181818182* |
| *Iron* | *0.675324675* | *0.519480519* | *0.346320346* | *0.277056277* | *1.818181818* |
| *Meat* | *0.272727273* | *0.090909091* | *0.303030303* | *0.242424242* | *0.909090909* |
| *Total* | *1.037748918* | *0.658874459* | *0.673593074* | *0.538874459* | *2.909090909* |

***Table 2. Aggregate value table*** (simple reproduction)

Solving the capital-allocation system (6.1a)–(6.1c) and then using $q_j = K_j/k_j$ gives $(q_1, q_2, q_3) = (8.109868, 0.355762, 0.757054)$.

For branch j, wages are calculated as $q_j(b_j^V + f_j^V + m_j^V) = q_j(\Lambda_1 v_1 \ell_j + 0 + \Lambda_3 v_3 \ell_j)$.

The corresponding aggregate value table is:

| | B | F | M | S | W | Wages |
|---|---|---|---|---|---|---|
| Bread | 0.727430628 | 0.393205745 | 0.196602872 | 0.157282298 | 1.474521543 | 0.167112 |
| Iron | 0.240255014 | 0.184811549 | 0.123207700 | 0.098566160 | 0.646840423 | 0.104727 |
| Meat | 0.206469385 | 0.068823128 | 0.229410428 | 0.183528342 | 0.688231284 | 0.194999 |
| Total | 1.174155027 | 0.646840423 | 0.549221000 | **0.439376800** | **2.809593250** | |

The magnitude of total capital $K_T$ can be fixed arbitrarily. In this example, $K_T$ is 2.370216 monetary units.[4]

***Table 3. Aggregate production-price table***

Prices are calculated using transformation coefficients ($x_1$, $x_2$, $x_3$) obtained by combining equations 5.9 and 5.10.

For branch j, wages are calculated as $q_j(x_1 b_j^V + x_2 f_j^V + x_3 m_j^V) = q_j(x_1 \Lambda_1 v_1 \ell_j + 0 + x_3 \Lambda_3 v_3 \ell_j)$.

| | $B^p$ | $F^p$ | $M^p$ | Π | $W^p$ | Wages |
|---|---|---|---|---|---|---|
| Bread | 0.784547437 | 0.393805927 | 0.163247423 | 0.248698072 | 1.590298859 | 0.168036 |
| Iron | 0.259119493 | 0.185093642 | 0.102304403 | 0.101310210 | 0.647827748 | 0.105305 |
| Meat | 0.222681065 | 0.068928179 | 0.190488881 | 0.089368518 | 0.571466643 | 0.196076 |
| Total | 1.266347995 | 0.647827748 | 0.456040706 | **0.439376800** | **2.809593250** | |

[4] In the numerical tables, the total capital advanced is normalised by setting $K_T = \Sigma k_j$. In the reference wage case, this gives $K_T = 2.370216$. This is the total advanced capital that would correspond to producing one unit of each commodity before imposing the reproduction closure conditions. It is only a normalisation of scale. Once Marx's aggregate equalities and the iron-reproduction condition are imposed, the same total capital must be reallocated across branches, and the resulting gross-output vector is no longer (1, 1, 1) but the vector q reported in Table 4b. Choosing another value of $K_T$, for example $K_T = 1000$, would simply rescale all $K_j$, $q_j$, $y_j$, and $h_j$ by factor 1000/2.370216, without changing the profit rate, the transformation coefficients, or the relative capital allocation.

***Table 4a. Profit rate, transformation coefficients, advanced capital and unit prices***

| **r** | $x_i$ | $K_i$ | $K_i^p$ | **Unit price** |
|---|---|---|---|---|
| **0.185374125** | 1.078518565 | 1.317239246 | 1.341600787 | 0.196094 |
| **0.185374125** | 1.001526382 | 0.548274263 | 0.546517538 | 1.820957 |
| **0.185374125** | 0.830340986 | 0.504702942 | 0.482098125 | 0.754855 |
| TOTAL | | **2.370216** | **2.370216** | |

In the reference case, $\Sigma W = \Sigma W^p = 2.809593250$ and $\Sigma S = \Sigma\Pi = 0.439376800$. The equality $K_T = K_T^p = 2.370216$ expresses the conservation of advanced capital in this circulating-capital reproduction scheme. The uniform profit rate is r = 0.185374125.

Physical quantities are obtained from equations (7a)-(7c). In the reference wage case, the net output of iron is zero, while bread and meat constitute the realised net outputs.

***Table 4b. Physical quantities.***

| **Commodity** | **Gross output $q_i$** | **Net output $y_i$** | **Surplus output $h_i$** |
|---|---|---|---|
| Bread | 8.109868 | 3.464445 | 1.652016 |
| Iron | 0.355762 | 0.000000 | 0.000000 |
| Meat | 0.757054 | 0.303947 | 0.152911 |

These quantities verify $W_i = \Lambda_i q_i$ for each commodity. For example, for bread, 1.474521543 = 0.181818182 × 8.109868. The fact that the net output of iron is zero does not imply that the iron sector produces no surplus-value. It only means that the entire physical output of iron is absorbed as productive input within the system.

**8.2 Zero-wage case: maximum profit rate**

$v = (0, 0, 0)^T$, $e \rightarrow +\infty$, $\alpha$ is not defined.

***Table 5. Unit decomposition with zero wages.***

| | b | f | m | s | Λ |
|---|---|---|---|---|---|
| Bread | 0.075151515 | 0.048484848 | 0.018181818 | 0.040000000 | 0.181818182 |
| Iron | 0.467532468 | 0.519480519 | 0.259740260 | 0.571428571 | 1.818181818 |
| Meat | 0.090909091 | 0.090909091 | 0.227272727 | 0.500000000 | 0.909090909 |
| Total | 0.633593074 | 0.658874459 | 0.505194805 | 1.111428571 | 2.909090909 |

***Table 6. Aggregate values with zero wages.***

| | B | F | M | S | W | Wages |
|---|---|---|---|---|---|---|
| Bread | 0.569569699 | 0.367464322 | 0.137799121 | 0.303158066 | 1.377991208 | 0 |
| Iron | 0.156686944 | 0.174096604 | 0.087048302 | 0.191506264 | 0.609338114 | 0 |
| Meat | 0.067777188 | 0.067777188 | 0.169442970 | 0.372774534 | 0.677771880 | 0 |
| Total | 0.794033831 | 0.609338114 | 0.394290393 | **0.867438864** | **2.665101202** | |

***Table 7. Production prices with zero wages.***

| | $B^p$ | $F^p$ | $M^p$ | Π | $W^p$ | Wages |
|---|---|---|---|---|---|---|
| Bread | 0.700396325 | 0.369983164 | 0.072598436 | 0.551529313 | 1.694507238 | 0 |
| Iron | 0.192676962 | 0.175289977 | 0.045860747 | 0.199687233 | 0.613514919 | 0 |
| Meat | 0.083345188 | 0.068241777 | 0.089269761 | 0.116222319 | 0.357079045 | 0 |
| Total | 0.976418475 | 0.613514919 | 0.207728944 | **0.867438864** | **2.665101202** | |

***Table 8a. Profit rate, transformation coefficients and capital totals with zero wages.***

| **r** | $x_i$ | $K_i$ | $K_i^p$ | **Unit price** |
|---|---|---|---|---|
| 0.482537152 | 1.229693795 | 1.074833142 | 1.142977926 | 0.223580690 |
| 0.482537152 | 1.006854658 | 0.417831850 | 0.413827686 | 1.830644832 |
| 0.482537152 | 0.526842519 | 0.304997346 | 0.240856726 | 0.478947744 |
| TOTAL | | **1.797662** | **1.797662** | |

With zero wages, the wage basket disappears and the profit rate reaches the maximum rate compatible with the technical system: r = 0.482537152. The aggregate equalities remain preserved: $\Sigma W = \Sigma W^p = 2.665101202$ and $\Sigma S = \Sigma \Pi = 0.867438864$.

With zero wages, $\tilde{A} = A$. The standard net output and the residual surplus output therefore coincide. The physical net outputs reported below are computed from A, as in all cases.

***Table 8b. Physical quantities in the zero-wage case.***

| Commodity | Gross output $q_i$ | Net output $y_i$ | Surplus output $h_i$ |
|---|---|---|---|
| Bread | 7.578952 | 3.211766 | 3.211766 |
| Iron | 0.335136 | 0.000000 | 0.000000 |
| Meat | 0.745549 | 0.311830 | 0.311830 |

**8.3 No surplus-value: maximum-wage cases**

In the zero-surplus case, the profit rate is zero and prices coincide with values; the capital-allocation system used above then becomes degenerate. Since $s_j=0$ for every branch j, one has $\Lambda_j=b_j+f_j+m_j=k_j$, and the uniform-profit-rate equations give $r=0$ and $x_1=x_2=x_3=1$. Branch profits therefore vanish identically:

$$\Pi_j=q_j[\Lambda_j-(b_j+f_j+m_j)]=0$$

for any scale of production. These branch equations do not determine the allocation of capital. The relevant closure is instead the zero-residual-surplus condition:

$$h=(I-\tilde{A})q=0.$$

In value form, this gives the three commodity-replacement equations:

$$q_1\Lambda_1=q_1b_1+q_2b_2+q_3b_3$$

$$q_2\Lambda_2=q_1f_1+q_2f_2+q_3f_3$$

$$q_3\Lambda_3=q_1m_1+q_2m_2+q_3m_3$$

Since their sum is identically satisfied when $\Lambda_j=b_j+f_j+m_j$, only two of these equations are independent. The relative vector q, or equivalently $K_j=q_jk_j$, is therefore determined up to scale; the absolute scale is fixed by the normalization $K_1+K_2+K_3=K_T$. In the numerical tables below, $K_T$ is set equal to $\Sigma_j k_j$.

The two following cases show that different wage-basket compositions can be compatible with r = 0, provided the aggregate reproduction conditions are respected.

**8.3.1 Maximum-wage case A: wage basket oriented toward meat (no bread)**

$\Lambda_3 v_3 = 1$, results in $v = (0, 0, 11/10)^T$, $\Lambda v = 1$, $e = 0$, $\alpha = 0$.

***Table 9. Unit decomposition in maximum-wage case A.***

| | b | f | m | s | Λ |
|---|---|---|---|---|---|
| Bread | 0.075151515 | 0.048484848 | 0.058181818 | 0 | 0.181818182 |
| Iron | 0.467532468 | 0.519480519 | 0.831168831 | 0 | 1.818181818 |
| Meat | 0.090909091 | 0.090909091 | 0.727272727 | 0 | 0.909090909 |
| Total | 0.633593074 | 0.658874459 | 1.616623377 | 0 | 2.909090909 |

***Table 10. Values and prices in case A.***

| Val = Pr | $B = B^p$ | $F = F^p$ | $M = M^p$ | $S = \Pi$ | $W = W^p$ | Wages |
|---|---|---|---|---|---|---|
| Bread | 0.219169537 | 0.141399701 | 0.169679641 | 0 | 0.530248879 | 0.116655 |
| Iron | 0.119773864 | 0.133082072 | 0.212931315 | 0 | 0.465787251 | 0.146390 |
| Meat | 0.191305478 | 0.191305478 | 1.530443824 | 0 | 1.913054779 | 1.052180 |
| Total | 0.530248879 | 0.465787251 | 1.913054779 | 0 | 2.909090909 | |

***Table 11a. Coefficients in case A.***

| r | xi | Ki=Kip |
|---|---|---|
| 0 | 1 | 0.530248879 |
| 0 | 1 | 0.465787251 |
| 0 | 1 | 1.913054779 |

In case A, surplus-value is zero. The residual surplus output vanishes, but the standard net output y remains positive for the wage good consumed by workers.

***Table 11b. Physical quantities in maximum-wage case A.***

| Commodity | Gross output $q_i$ | Net output $y_i$ | Surplus output $h_i$ |
|---|---|---|---|
| Bread | 2.916369 | 0.000000 | 0.000000 |
| Iron | 0.256183 | 0.000000 | 0.000000 |
| Meat | 2.104360 | 1.446748 | 0.000000 |

### 8.3.2 Maximum-wage case B: wage basket oriented toward bread (no meat)

$\Lambda_1 v_1 = 1$ results in $v = (11/2, 0, 0)^T$, $\Lambda v = 1$, $e = 0$, $\alpha = 1$.

***Table 12. Unit decomposition in maximum-wage case B.***

| | b | f | m | s | Λ |
|---|---|---|---|---|---|
| Bread | 0.115151515 | 0.048484848 | 0.018181818 | 0 | 0.181818182 |
| Iron | 1.038961039 | 0.519480519 | 0.259740260 | 0 | 1.818181818 |
| Meat | 0.590909091 | 0.090909091 | 0.227272727 | 0 | 0.909090909 |
| Total | 1.745021645 | 0.658874459 | 0.505194805 | 0 | 2.909090909 |

***Table 13. Values and prices in case B.***

| Val = Pr | $B = B^p$ | $F = F^p$ | $M = M^p$ | $S = \Pi$ | $W = W^p$ | Wages |
|---|---|---|---|---|---|---|
| Bread | 1.142386492 | 0.481004839 | 0.180376815 | 0 | 1.803768146 | 0.396829 |
| Iron | 0.415113765 | 0.207556883 | 0.103778441 | 0 | 0.726449089 | 0.228313 |
| Meat | 0.246267888 | 0.037887367 | 0.094718419 | 0 | 0.378873674 | 0.208381 |
| Total | 1.803768146 | 0.726449089 | 0.378873674 | 0 | 2.909090909 | |

***Table 14a. Coefficients in case B.***

| r | xi | Ki=Kip |
|---|---|---|
| 0 | 1 | 1.803768146 |
| 0 | 1 | 0.726449089 |
| 0 | 1 | 0.378873674 |

In case B as well, surplus-value is zero and the residual surplus output vanishes, while the standard net output y remains positive for the wage good consumed by workers.

***Table 14b. Physical quantities in maximum-wage case B.***

| Commodity | Gross output $q_i$ | Net output $y_i$ | Surplus output $h_i$ |
|---|---|---|---|
| Bread | 9.920725 | 4.584371 | 0.000000 |
| Iron | 0.399547 | 0.000000 | 0.000000 |
| Meat | 0.416761 | 0.000000 | 0.000000 |

Because $S = \Pi = 0$ in these cases, $x_i = 1$ and the transformation is irrelevant. The example is useful because it shows that the wage basket affects the technical-price system through the augmented matrix, but without surplus-value no redistribution of surplus-value can occur.

## 9. Discussion

### 9.1 Counting the degrees of freedom

The degree of freedom follows from an explicit count for a general n-branch model. The unknowns are the n transformation coefficients, the uniform profit rate, and the n advanced capitals: 2n + 1 in all. The equations are the n uniform-profit conditions, Marx's two aggregate equalities, and the fixed total $\Sigma K_j = K_T$: n + 3 in all. The system therefore carries n

− 2 degrees of freedom. For n = 2 the allocation is uniquely determined; for n = 3 the admissible allocations form a one-parameter family, which a further condition selects — here the reproduction condition (6.1c).

This count locates the source of the classical impossibility results. Von Bortkiewicz, and those who follow him, treat the sectoral capitals as data. Doing so removes n unknowns, leaving n + 1 unknowns for n + 2 equations, and the resulting overdetermination is read as an inconsistency in Marx. But the overdetermination follows from taking the sectoral capitals as given, not from the theory: it freezes the very magnitudes whose determination is at issue. Treated as unknowns, they leave a system that is exactly determined for two branches and admits a one-parameter family of solutions for three.

**9.2 The role of the stationary assumption**

The stationary assumption adopted here means that the technical coefficients, direct labour requirements, wage basket, labour-values and production prices are held fixed over the reproduction period considered. It should not be interpreted as a claim that capitalist economies are empirically stationary. Rather, it provides a benchmark for examining the internal coherence of the transformation of values into prices of production under a given socio-technical configuration (on this use of abstraction, cf. Rubin, 1972 [1928]). The capital allocations derived in this paper therefore describe admissible stationary configurations, not the adjustment paths through which capital moves between branches. In a dynamic setting, capital mobility may alter production conditions through changes in scale, innovation and the adoption of new techniques, thereby making technical coefficients and labour requirements endogenous. Such dynamic interactions between capital mobility, technical change, price formation, disequilibrium and devaluation lie beyond the scope of the present paper and are left for subsequent research.

### 9.3 Technical productivity is not social realisation

The Hawkins-Simon condition ensures that the technical system can produce at least its inputs. It does not ensure that commodities are saleable at prices of production.

Three levels must therefore be distinguished. The Hawkins-Simon condition concerns technical reproducibility: it says whether the system can physically reproduce its inputs. The reproduction constraints concern material coherence: they specify which commodities must be replaced as inputs and which commodities may appear as net outputs. Market validation concerns realised value: it determines whether the residual outputs can actually count as socially validated value. In the stationary benchmark considered here, the effective-demand constraint is satisfied by assuming that the produced residual outputs coincide with effective social demand. This is why commodities with a final outlet limited to the technical requirements of production, such as iron in the present example, must have zero net output, whereas wage goods and surplus-consumption goods may have positive net outputs.

A commodity that would be produced but would lack effective demand would have no realised social value. Marx's fundamental equalities apply to realised values, not to merely produced and unsold commodities. This point has been made for the two-sector model developed in Ankri (in press). It highlights a limitation of formalisms in which the net-output vector enters as a given physical datum or as a normalising magnitude. In a capitalist economy, the vector of net outputs cannot be treated as an arbitrary physical datum independently of social validation. Morishima, for example, explicitly refers to the production of "given amounts of the final net outputs" in his optimum labour-value formulation (Morishima, 1973), while Duménil and Foley use the net product as the numéraire in their treatment of the transformation problem (Duménil and Foley, 2006)[5]. This does not imply that these authors deny the role of social demand; rather, the point is that effective social demand is not introduced as an independent constraint determining the realised price–quantity configuration.

The net-output vector is constrained by the interaction between technical reproduction and effective social demand. A commodity may be technically producible, but its value is realised only if it can be sold. The net-output vector must therefore be understood as an outcome of the economy as a whole, not as a freely chosen starting point. Treating such a vector as independent of demand would correspond more naturally to a planned economy than to a capitalist market economy. Even a planned economy would have to solve the problem of adjusting production to social needs, but it would not do so through the equalisation of profit rates.

### 9.4 Wage-basket composition

The wage basket enters the augmented matrix $\tilde{A} = A + v\ell$ in equation (5.1). Therefore, two wage baskets with the same value but different commodity compositions - for instance, different values of $\alpha$ in equation (3) - may lead to different uniform profit rates, depending on whether the goods in the wage basket correspond to more or less capital-intensive branches. This does not contradict Marx's theory. It means that the value of labour-power fixes the rate of exploitation, while the commodity composition through which that value is reproduced may influence the intersectoral structure of prices and profitability. On this point the present result differs from the conclusion drawn by Duménil (1982) and Lipietz (1982), for whom prices of production should be determined independently of workers' consumption.

[5] Morishima writes that "the optimum labour theory of value determines the methods of production which minimize the amount of labour needed to produce given amounts of the final net outputs" (Morishima, 1973, p. 195). Duménil and Foley define their numéraire by reference to the net product: "the numéraire is the net product divided by the total number of hours expended: $N = y/(l'x)$" (Duménil and Foley, 2006, p. 13).

### 9.5 Capital-intensive branches and price deviations

The model clarifies why capital-intensive branches may attract capital even though they produce less surplus-value locally per unit of capital advanced. They may sell above value through the transformation process and thereby appropriate a share of surplus-value produced elsewhere. This is not a refutation of the labour theory of value but its price-form expression. At the aggregate level, the redistributed profit remains bounded by realised surplus-value.

### 9.6 Disequilibrium, devaluation and temporal feedback

The restriction adopted in this paper should not be confused with the claim that capitalist production is normally adjusted perfectly to effective demand. If production exceeds socially validated demand, part of the value potentially initiated in production is not realised. The market does not create value independently of production; rather, it retroactively validates or invalidates the potential social value initiated in production. When commodities are produced beyond effective social demand, the corresponding labour is not fully recognised as socially necessary. Part of the potential value is therefore devalued or destroyed. In such disequilibrium cases, market feedback modifies realised values, and the prices of inputs and outputs need not coincide. This opens the way to a temporal analysis of reproduction and crisis, as discussed in Ankri (in press) and in Ankri and Marcaggi (2022), and connects with concerns emphasized by temporal approaches to Marx's value theory, notably Freeman (1995) and Kliman (2007). The present paper deliberately abstracts from this case in order to examine the coherent stationary benchmark in which the whole realised social product is validated through sale.

This also clarifies the relation between the present approach, the New Interpretation and temporal approaches. Unlike the New Interpretation, which uses the net product to define the monetary expression of labour or the numéraire, the present paper does not take the realised net product as an independent normalising datum. It treats the realised net product as part of the price–quantity configuration to be constrained by reproduction and effective social demand. Conversely, the present paper shares with temporal approaches the idea that disequilibrium may involve a discrepancy between input prices and output prices and the devaluation of unrealised commodities.

## 10. Conclusion

The three-branch input-output model confirms the central intuition of the two-sector law while showing why a more complex formalism is required beyond two branches. A Morishima-type formalism provides the appropriate linear-algebraic framework to compute labour-values from the technical matrix (A). Defining the augmented matrix $\tilde{A}$ (A with incorporation of the wage basket) further enables to compute the uniform-profit-rate price system. Yet this is not a complete transformation unless the model also imposes the aggregate conservation constraints expressed in equations (5.11) and (5.12), together with the relevant reproduction condition. Once these conditions are made explicit, the apparent contradiction between values and prices disappears. Production prices redistribute surplus-value; they do not create it. Capital allocation is not arbitrary; it is constrained by the reproduction of the system. The result is a coherent stationary benchmark, based on full realisation of the social product, from which the dynamic interaction between capital mobility, technical change, disequilibrium, devaluation and temporal price discrepancies can be studied.

## 11. Appendix. Generalization to n branches

Let A be an n × n non-negative productive input-output matrix, ℓ the row vector of direct labour requirements, and v the column vector of the real wage basket. The vector of labour-values is

$$\Lambda = \ell(I - A)^{-1}.$$

The augmented matrix incorporating wage-goods consumption is

$$\tilde{A} = A + v\ell,$$

with entries $\tilde{a}_{ij} = a_{ij} + v_i \ell_j$. Using a row-vector convention for prices in this appendix, prices of production satisfy

$$p = (1 + r)p\tilde{A},$$

$$p\tilde{A} = \lambda p, \quad r = 1/\lambda - 1.$$

Although the eigenvector is mathematically determined only up to scale, the economically relevant scale is fixed by the aggregate equality between total price and total realised value:

$$\Sigma_i \, p_i q_i = \Sigma_i \, \Lambda_i q_i.$$

The physical-quantity counterpart of the reproduction argument in section 7 can be expressed by distinguishing the standard net-output vector from the residual surplus-output vector. Let y denote the standard net-output vector after replacement of the technical input requirements recorded in A:

$$q = Aq + y, \quad y = (I - A)q, \quad q = (I - A)^{-1}y \qquad \text{(A1)}$$

Since wage-goods consumption is incorporated in Ã, the residual surplus-output vector is instead

$$h = (I - \tilde{A})q = y - v(\ell q). \tag{A2}$$

The connection between h and total surplus-value follows directly from the value equation. Since $\Lambda = \ell(I - A)^{-1}$, we have

$$\Lambda(I - A) = \ell.$$

For the standard net-output vector $y = (I - A)q$, it follows that

$$\Lambda y = \ell q. \tag{A3}$$

Thus the value of the standard net product is equal to the total new value created by living labour. Since $h = y - v(\ell q)$, we obtain

$$\Lambda h = \Lambda y - \Lambda v(\ell q) \tag{A4}$$

$$= \ell q - \Lambda v(\ell q) = (1 - \Lambda v)\ell q.$$

But total surplus-value is

$$\Sigma_j S_j = \Sigma_j q_j[\ell_j - (\Lambda v)\ell_j] = (1 - \Lambda v)\ell q. \tag{A5}$$

Therefore,

$$\Lambda h = \Sigma_j S_j. \tag{A6}$$

The vector h is therefore the physical residual surplus-output vector whose value expression is equal to total surplus-value. In the zero-surplus limiting case, $\Lambda h = 0$. If $h_i \geq 0$ for all i and $\Lambda_i > 0$ for all i, this implies $h_i = 0$ for every commodity. With positive surplus-value, the components of h need not vanish; they represent residual surplus commodities validated by effective demand.

When the production-price scale is fixed by Marx's aggregate equalities, the price expression of the same residual surplus output is equal to total profit:

$$\Sigma_i x_i \Lambda_i h_i = \Sigma_j \Pi_j = \Sigma_j S_j. \tag{A7}$$

However, for any commodity that is neither a wage good nor an object of capitalist consumption, the relevant reproduction condition is $h_i = 0$; if $v_i = 0$, this is equivalent to $y_i = 0$. In the bread-iron-meat example, this is precisely the role of the iron condition (7d), equivalently equation (6.4). In an n-branch model, the number and nature of such constraints depend on which commodities are consumed as wage goods, which are used only as means of production, whether capitalist consumption is represented explicitly, and whether fixed capital is introduced. The general conclusion is unchanged: technical productivity, value conservation and effective demand jointly determine the admissible transformation.

Let F denote the set of commodities that can absorb socially validated final demand, and let P denote the set of commodities used only as means of production. For commodities $i \in P$, the relevant simple-reproduction condition is $h_i = 0$, or equivalently $y_i = 0$ when $v_i = 0$. If $|P|$ such conditions are independent, then, in the positive-surplus case, the allocation of n sectoral capital scales is constrained by $|P|$ zero-output conditions, by the aggregate profit–surplus-value equality, and by one normalisation such as $\Sigma_j K_j = K^T$. Hence the expected dimension of the admissible allocation set is $n - (|P| + 2) = |F| - 2$, provided the corresponding system has full rank. In the three-branch example, $|F| = 2$, so the admissible allocation is generically unique once the normalisation $\Sigma_j K_j = K_T$ is imposed. If $|F| > 2$, the transformation does not select a unique realised price–quantity configuration. It defines an admissible space constrained by technical productivity, Marx's aggregate invariants and effective demand; further behavioural or institutional assumptions concerning capitalist consumption, accumulation, competition or market power would then be required to select a particular point in that space. In the zero-surplus limiting case, the profit–surplus-value equation becomes degenerate, and the appropriate closure is instead the zero-residual-surplus condition $h = 0$, as shown in section 8.3.